\documentstyle[11pt,twoside]{article}

\begin{document}

\newcommand{\dd}{\,{\rm d}}
\newcommand{\ie}{{\it i.e.},\,}
\newcommand{\etal}{{\it et al.\ }}
\newcommand{\eg}{{\it e.g.},\,}
\newcommand{\cf}{{\it cf.\ }}
\newcommand{\vs}{{\it vs.\ }}
\newcommand{\zdot}{\makebox[0pt][l]{.}}
\newcommand{\up}[1]{\ifmmode^{\rm #1}\else$^{\rm #1}$\fi}
\newcommand{\dn}[1]{\ifmmode_{\rm #1}\else$_{\rm #1}$\fi}
\newcommand{\upd}{\up{d}}
\newcommand{\uph}{\up{h}}
\newcommand{\upm}{\up{m}}
\newcommand{\ups}{\up{s}}
\newcommand{\arcd}{\ifmmode^{\circ}\else$^{\circ}$\fi}
\newcommand{\arcm}{\ifmmode{'}\else$'$\fi}
\newcommand{\arcs}{\ifmmode{''}\else$''$\fi}
\newcommand{\MS}{{\rm M}\ifmmode_{\odot}\else$_{\odot}$\fi}
\newcommand{\RS}{{\rm R}\ifmmode_{\odot}\else$_{\odot}$\fi}
\newcommand{\LS}{{\rm L}\ifmmode_{\odot}\else$_{\odot}$\fi}

\newcommand{\Abstract}[2]{{\footnotesize\begin{center}ABSTRACT\end{center}
\vspace{1mm}\par#1\par
\noindent
{\bf Key words:~~}{\it #2}}}

\newcommand{\TabCap}[2]{\begin{center}\parbox[t]{#1}{\begin{center}
  \small {\spaceskip 2pt plus 1pt minus 1pt T a b l e}
  \refstepcounter{table}\thetable \\[2mm]
  \footnotesize #2 \end{center}}\end{center}}

\newcommand{\TableSep}[2]{\begin{table}[p]\vspace{#1}
\TabCap{#2}\end{table}}

\newcommand{\TableFont}{\footnotesize}
\newcommand{\TableFontIt}{\ttit}
\newcommand{\SetTableFont}[1]{\renewcommand{\TableFont}{#1}}

\newcommand{\MakeTable}[4]{\begin{table}[htb]\TabCap{#2}{#3}
  \begin{center} \TableFont \begin{tabular}{#1} #4 
  \end{tabular}\end{center}\end{table}}

\newcommand{\MakeTableSep}[4]{\begin{table}[p]\TabCap{#2}{#3}
  \begin{center} \TableFont \begin{tabular}{#1} #4 
  \end{tabular}\end{center}\end{table}}

\newenvironment{references}%
{
\footnotesize \frenchspacing
\renewcommand{\thesection}{}
\renewcommand{\in}{{\rm in }}
\renewcommand{\AA}{Astron.\ Astrophys.}
\newcommand{\AAS}{Astron.~Astrophys.~Suppl.~Ser.}
\newcommand{\ApJ}{Astrophys.\ J.}
\newcommand{\ApJS}{Astrophys.\ J.~Suppl.~Ser.}
\newcommand{\ApJL}{Astrophys.\ J.~Letters}
\newcommand{\AJ}{Astron.\ J.}
\newcommand{\IBVS}{IBVS}
\newcommand{\PASP}{P.A.S.P.}
\newcommand{\Acta}{Acta Astron.}
\newcommand{\MNRAS}{MNRAS}
\renewcommand{\and}{{\rm and }}
\section{{\rm REFERENCES}}
\sloppy \hyphenpenalty10000
\begin{list}{}{\leftmargin1cm\listparindent-1cm
\itemindent\listparindent\parsep0pt\itemsep0pt}}%
{\end{list}\vspace{2mm}}

\def\TYLDA{~}
\newlength{\DW}
\settowidth{\DW}{0}
\newcommand{\dw}{\hspace{\DW}}

\newcommand{\refitem}[5]{\item[]{#1} #2%
\def\REFARG{#3}\ifx\REFARG\TYLDA\else, {\it#3}\fi
\def\REFARG{#4}\ifx\REFARG\TYLDA\else, {\bf#4}\fi
\def\REFARG{#5}\ifx\REFARG\TYLDA\else, {#5}\fi.}

\newcommand{\Section}[1]{\section{#1}}
\newcommand{\Subsection}[1]{\subsection{#1}}
\newcommand{\Acknow}[1]{\par\vspace{5mm}{\bf Acknowledgments.} #1}
\pagestyle{myheadings}


\def\thefootnote{\fnsymbol{footnote}}

\begin{center}
{\Large\bf Advection Dominated Accretion Flows.\\}
\vskip3pt
{\Large\bf A Toy Disk Model$^*$}
\vskip1cm
{\bf B~o~h~d~a~n~~P~a~c~z~y~{\'n}~s~k~i}
\vskip6mm
{Princeton University Observatory, Princeton, NJ 08544-1001, USA\\
e-mail: bp@astro.princeton.edu}
\end{center}
\vskip1cm
\Abstract{
A toy model of a disk undergoing steady state accretion onto a black hole 
is presented.  The disk is in
a hydrostatic equilibrium for all radii $ r > r_{in} $, with the inner disk
radius located between the marginally stable and marginally bound orbits:
$ r_{ms} > r_{in} > r_{mb} $.  Matter flows from the disk through a narrow
cusp at $ r_{ms} $ and falls freely into the black hole, carrying with it
no thermal energy.  At radii larger than $ r_{out} $ the disk is assumed to
radiate away all locally generated heat, and therefore the disk is 
geometrically thin for $ r > r_{out} $.  We assume that no heat generated
in the inner disk, with $ r_{out} > r > r_{in} $ can be radiated away, i.e.
the disk is 100\% advective, and it becomes geometrically thick in this
range of radii.  All enthalpy of the thick disk is used up to press
the inner disk radius towards the marginally bound orbit, and to lower the
efficiency of conversion of accreted mass into radiation generated
only for $ r > r_{out} $, by assumption.

Conservation laws of mass, angular momentum and energy make it possible
to calculate the inner thick disk radius $ r_{in} $ for any specified value
of its outer radius $ r_{out} $.  As the nature of disk viscosity is not
known there is some freedom in choosing the shape of the thick disk, subject
to several general conditions, which include the hydrostatic equilibrium 
everywhere for $ r > r_{in} $.  The main purpose of this toy model is to
emphasize the effect the disk thickness has on lowering the energetic 
efficiency of a black hole accretion.
} {accretion, accretion disks -- black hole physics}

\Section{Introduction}

Following the publication of the first paper on the advection dominated
accretion flows by Narayan and Yi (1994) over one hundred papers appeared
on this subject, with ever more sophisticated physics and ever more
sophisticated treatment of geometry (cf. Gammie and Popham 1998, and
references therein).  However, there is no fully two dimensional treatment
\footnote{published 1998, Acta Astronomica, 48, 667.}
published so far, and there is no clear link between the recent papers
and those written about two decades ago on the theory of thick disks
(e.g. Jaroszy\'nski, Abramowicz and Paczy\'nski 1980, Paczy\'nski and Wiita 
1980, and references therein).  The purpose
of this paper is to present a toy model of a disk accreting onto a black hole
but not radiating, i.e. advection dominated, presented in the spirit of the
early 1980s, which seems to be simpler and more transparent than the spirit
dominating the late 1990s.

\Section{Thin Pseudo-Newtonian Disk}

I shall use pseudo-Newtonian gravitational potential (Paczy\'nski and Wiita 
1980):
$$
\Psi = - { GM \over R - R_g } , 
\hskip 1.0cm \Psi ' \equiv { d \Psi \over dR } = { GM \over (R-R_g)^2 } ,
\hskip 1.0cm R_g \equiv { 2GM \over c^2 } ,
\eqno(1)
$$
which has the property that a test particle has a marginally stable orbit at 
$ R_{ms} = 3 R_g $, and marginally bound orbit at $ R_{mb} = 2 R_g $, just as 
it is the case with particles orbiting Schwarzschild black hole.  $ R $ is
the spherical radius.  Throughout this paper cylindrical coordinates $ (r,z) $
will be used, with $ R = \left( r^2 + z^2 \right)^{1/2} $.

In the following we consider a thin accretion disk fully supported against
gravity by the centrifugal acceleration.  The disk is in the equatorial plane
of the coordinate system.  It follows the motion of test particles
on circular orbits, with the rotational velocity $ v(r) $, angular velocity 
$ \Omega (r) $, specific angular momentum $ j(r) $, and total specific energy 
$ e(r) $ given as
$$
v = \left( r \Psi ' \right) ^{1/2} =
\left( { GM \over r } \right) ^{1/2} \left[ { r \over r - r_g } \right] ,
\eqno(2a)
$$
$$
\Omega = { v \over r } =  \left( { \Psi ' \over r } \right) ^{1/2} =
\left( { GM \over r^3 } \right) ^{1/2} \left[ { r \over r - r_g } \right] ,
\eqno(2b)
$$
$$
j = vr = \left( r^3 \Psi ' \right) ^{1/2} =
\left( { GMr } \right) ^{1/2} \left[ { r \over r - r_g } \right] ,
\eqno(2c)
$$
$$
e = \Psi + { v^2 \over 2 } = \left( - { GM \over 2 r } \right)
\left[ { ( r - 2 r_g ) r \over ( r - r_g )^2 } \right] ,
\eqno(2d)
$$
where we adopted $ r_g \equiv R_g $.  It follows that
$$
{ de \over dr } = \Omega { dj \over dr } .
\eqno(3)
$$

The inner edge of a thin accretion disk is at 
$$ 
r_{in} = r_{ms} = 3 r_g ,
\eqno(4)
$$
where the binding energy and the specific angular momentum reach their minima:
$$
e_{ms} = - { c^2 \over 16 } ,
\eqno(5)
$$
$$
j_{ms} = 1.5 \times 6^{1/2} \times { GM \over c } \approx 3.674 ~
{ GM \over c } .
\eqno(6)
$$
The matter falls freely into the black hole once it crossed the $ r_{ms} $,
conserving its angular momentum and total energy.

The total thin disk luminosity is given with the formula:
$$
L_d = \dot M e_{in} = \dot M e_{ms} =
\left( - \dot M \right) ~ { c^2 \over 16 } ,
\hskip 1.5cm {\rm (thin ~ disk),}
\eqno(7)
$$
where I adopt a convention that $ \dot M < 0 $ for the accretion flow.

In a steady state accretion of a thin disk the equations of
mass and angular momentum conservation may be written as
$$
\dot M = const.
\eqno(8a)
$$
$$
\dot J = \dot M j + g = const.
\eqno(8b)
$$
where $ g $ is the torque acting between two adjacent rings in the disk.
With no torque at the inner edge of the disk at $ r_{in} = r_{ms} $ we have
$$
g = \left( - \dot M \right) \left( j - j_{in} \right) =
\left( - \dot M \right) \left( j - j_{ms} \right) .
\eqno(9)
$$

The rate of energy flow in the disk is given as
$$
\dot E = \dot M e + g \Omega .
\eqno(10)
$$
This is not constant, as the disk must radiate energy dissipated by viscous 
stresses in order to remain thin.  The energy balance equation may be
written as
$$
2 F \times 2 \pi r = - { d \dot E \over dr } = 
g \left( - { d \Omega \over dr } \right) =
\left( - \dot M \right) \left( j - j_{in} \right)
\left( - { d \Omega \over dr } \right) .
\eqno(11)
$$
where $ F $ is the energy radiated per unit disk area from each of its
two surfaces.  The eq. (11) may be integrated from $ r_{ms} $ to
infinity to obtain the same result as that given with the eq. (7).

Note, that disk viscosity was not specified.  All
we needed were the conservation laws and the assumption of a steady state
accretion of a geometrically thin disk.  Small geometrical thickness implied
that no significant internal energy could be stored within the disk.

\Section{Thick Pseudo-Newtonian Disk}

I shall consider now geometrically thick, axisymmetric disk, with the
surface given with the relation $ z_s (r) $, where $ (r,z) $ are the
two cylindrical coordinates, and we have
$$
R = \left( r^2 + z^2 \right) ^{1/2} .
\eqno(12)
$$
The disk is assumed to be in a hydrostatic equilibrium.  This implies that
the three vectors have to balance at its surface: gravitational acceleration,
centrifugal acceleration, and the pressure gradient divided by gas density.
This implies that angular velocity $ \Omega _s $ at the disk surface is given 
as (cf. Jaroszy\'nski et al. 1980, Paczy\'nski and Wiita 1980):
$$
\Omega _s^2 = \left( { \Psi_s ' \over R_s } \right) 
\left( { z_s \over r } { dz_s \over dr } + 1 \right) ,
\eqno(13a)
$$
which is equivalent to 
$$
{ de_s \over dr } = \Omega _s { dj_s \over dr } ,
\eqno(13b)
$$
where subscript `s' indicates that the corresponding quantities
are defined at the disk surface.

Thick disk has a large radial pressure gradient, and it remains in a
hydrostatic equilibrium for $ r > r_{in} $, and it has a cusp at $ r_{in} $.
The matter falls freely into the black hole inwards of $ r_{in} $, which
is located between the marginally stable and marginally bound orbits, i.e.
$$
r_{mb} < r_{in} < r_{ms}
\eqno(14)
$$
where a no torque inner boundary condition is applied.  I shall not analyze
the transition from a sub-sonic accretion flow at $ r > r_{in} $ to a supersonic
flow at $ r < r_{in} $ (cf. Loska 1982).

The total disk luminosity is given with the formula similar to eq. (7):
$$
L_d = \dot M e_{in} =
\left( - \dot M \right) { GM ( r_{in} - 2 r_g )  \over 
2 ( r_{in} - r_g )^2 } ,
\hskip 1.5cm {\rm (thick ~ disk).}
\eqno(15)
$$
The specific binding energy at $ r_{in} $ has the range
$$
0 > e_{in} > - { c^2 \over 16 } \hskip 1.0cm {\rm for} \hskip 1.0cm
r_{mb} < r_{in} < r_{ms} .
\eqno(16)
$$
Note, that while the total luminosity of a thick disk may be larger than
that of a thin disk, the energy radiated per each gram of matter accreted
into the black hole is smaller in the thick disk case.

\Section{Thick Advection Dominated Disk}

A thin accretion disk has to radiate in order to remain thin.  If the disk
is not radiating away the energy which is generated by viscous stresses
then it must be thick in order to accommodate this energy in its interior.
Let us consider a disk which is thin for $ r > r_{out} $, but is
thick for $ r_{out} > r > r_{in} $.  Note, that the disk is also thin
at its cusp, i.e. at $ r = r_{in} $.  
Our new disk radiates energy only where it is thin, i.e. for
$ r > r_{out} $, where the disk surface brightness $ F $
is given with the eq. (11), and the total disk luminosity is given by the
formula similar to the eq. (7) and (15):
$$
L_d = \dot M e_{in} .
\eqno(17)
$$
All this luminosity is radiated by the thin disk at $ r > r_{out} $, 
and none is radiated by the thick disk at $ r_{out} > r > r_{in} $, by 
assumption.

The structure of the thick disk, extending from  $ r_{in} $ to $ r_{out} $
is of interest for us, as by assumption the disk is 100\% advection
dominated in this range of radii, i.e. the mass, angular momentum, and
energy entering it at $ r_{out} $ must come out of it at $ r_{in} $:
$$
\dot M _{in} = \dot M_{out} .
\eqno(18a)
$$
$$
\dot J _{in} = \left( \dot M j + g \right) _{in} = 
\left( \dot M j + g \right) _{out} = \dot J _{out} ,
\eqno(18b)
$$
$$
\dot E _{in} = \left( \dot M e + g \Omega \right) _{in} = 
\left( \dot M e + g \Omega \right) _{out} = \dot E _{out} 
\eqno(18c)
$$
with $ j $ and $ \Omega $ having their `Keplerian' values (cf. eqs. 2)
at the two ends, as the disk is thin at both ends, and there is no
torque at $ r_{in} $, i.e. $ g_{in} = 0 $.

\begin{figure}[t]
\vspace{8.0cm}
\includegraphics{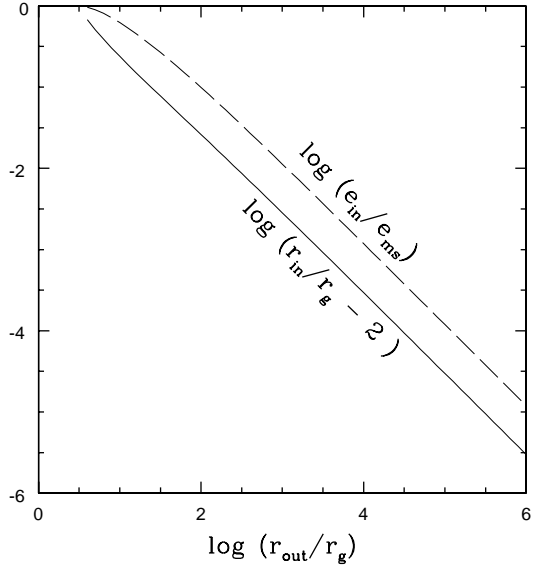}
\caption{
The variation of the inner thick disk radius $ r_{in} $ and the corresponding
binding energy $ e_{in} $ is shown as a function of the outer thick disk
radius $ r_{out} $.  $ r_g = 2GM/c^2 $ is the black hole radius, and 
$ e_{ms} = -c^2/16 $ is binding energy of a test particle with a unit mass 
orbiting the black hole at the marginally stable orbit with radius 
$ r_{ms} = 3 r_g $.
}
\end{figure}

The torque at $ r_{out} $ follows from the conservation of angular momentum
(18b):
$$
g_{out} = \left( - \dot M \right) \left( j_{out} - j_{in} \right) ,
\eqno(19)
$$
which, together with the energy conservation law give us:
$$
e_{out} - e_{in} = \left( j_{out} - j_{in} \right) \Omega _{out} .
\eqno(20)
$$
Note, that all quantities in the eq. (20) are unique functions of either
$ r_{in} $ or $ r_{out} $ (cf. eqs. 2), and therefore the eq. (20) gives the
relation between $ r_{in} $ and $ r_{out} $.  If the outer radius of the
advection dominated disk is specified, then the inner radius of the disk can 
be calculated with the eq. (20), and the eq. (17) gives the total luminosity
of the thin disk which is assumed to extend from $ r_{out} $ to infinity.
The relation between $ r_{out} $ and $ r_{in} $ is shown in Figure 1.
The larger is $ r_{out} $, the smaller is $ r_{in} $, and the less energy
is radiated per one gram of accreted matter.

The variation of the inner thick disk radius $ r_{in} $, and the binding energy
at the inner thick disk radius $ e_{in} $ with the outer thick disk radius 
$ r_{out} $ is shown in Fig. 1.  Notice, that for very large $ r_{out} $
we have asymptotically
$$
{ r_{in} \over r_g } \approx 2 +  3 { r_g \over r_{out} } , \hskip 1.0cm
{ e_{in} \over e_{ms} } \approx 12 { r_g \over r_{out} } , \hskip 1.0cm
{\rm for} \hskip 1.0cm  { r_{out} \over r_g } \gg 1 .
\eqno(21)
$$
We also have
$$
e_{out} \approx - { c^2 \over 4 } { r_g \over r_{out} } , \hskip 0.8cm
e_{in} \approx 3 e_{out} \approx - { 3 c^2 \over 4 } { r_g \over r_{out} } , 
\hskip 0.8cm {\rm for} \hskip 0.8cm  { r_{out} \over r_g } \gg 1 .
\eqno(22)
$$

So far we used only conservation laws to constrain our thick advective
disk.  If we want to find the disk shape we must make some additional
assumptions.  There are two general inequalities which must be satisfied
by the matter at the disk surface (cf. Jaroszy\'nski et al. 1980, Paczy\'nski
and Wiita 1980):
$$
{ d j_s \over dr } > 0 , \hskip 1.5cm
{d \Omega _s \over dr } < 0 ,
\eqno(23)
$$
supplemented with the condition of hydrostatic equilibrium at the thick disk
surface, as expressed with the eq. (13b), and the conditions that the disk
must be geometrically thin at $ r_{in} $ and $ r_{out} $.

There is a lot of freedom in choosing thick disk structure that satisfies
all the conditions listed in the previous paragraph.  For our toy model
we adopted
$$
j_s = j_{in} \left[ 1 + b \left( { x_s \over x_{in} } - 1 \right) ^a
+ b \left( { x_s \over x_{in} } - 1 \right) ^{3a} \right] ^{1.5/a} ,
\eqno(24)
$$
and
$$
e_s = e_{in} + \int _{r_{in}}^{r_s} \Omega _s { d j_s \over dr } dr =
e_{in} + \int _{r_{in}}^{r_s}  { 1 \over 2r^2 } { d j_s^2 \over dr } dr .
\eqno(25)
$$
Thin disk conditions at $ r_{in} $ are satisfied automatically with
the eqs. (24) and (25), and we have
$$
j_{in} = { \left( { GMr_{in}^3 } \right) ^{1/2} \over r_{in} - r_g } ,
\hskip 1.5cm
e_{in} = - { GM ( r_{in} - 2 r_g ) \over 2 ( r_{in} - r_g )^2 } .
\eqno(26)
$$
The parameters $ a $ and $ b $ have to be adjusted so that thin disk conditions
are satisfied at $ r_{out} $, where the eqs. (24) and (25) must give
$$
j_{out} = { \left( { GMr_{out}^3 } \right) ^{1/2} \over r_{out} - r_g } ,
\hskip 1.5cm
e_{out} = - { GM ( r_{out} - 2 r_g ) \over 2 ( r_{out} - r_g )^2 } .
\eqno(27)
$$

\begin{figure}[t]
\vspace{8.0cm}
\includegraphics{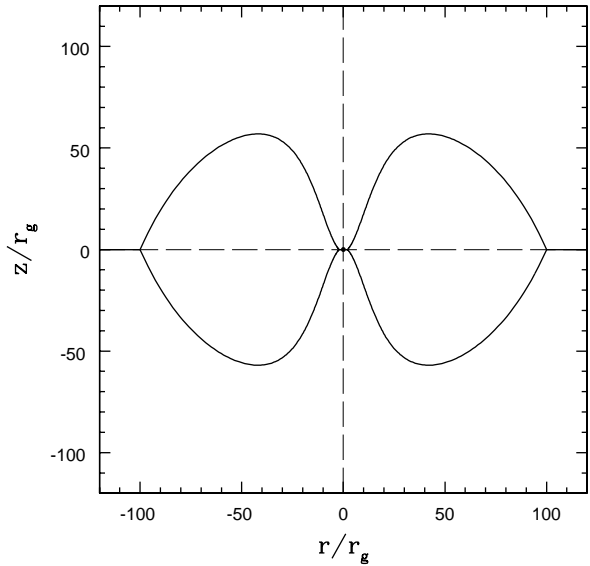}
\caption{
A cross section of the toy disk model accreting into a black hole.  All energy 
generated inwards of $ r_{out} = 100 ~ r_g $ is assumed to be
advected with the accretion flow, and none is radiated away.  At large
distance, $ r > r_{out} $, all energy generated by viscosity is assumed
to be radiated away, hence the disk is geometrically thin for $ r > r_{out} $.
The inability to radiate away energy for $ r < r_{out} $ forces the disk
to become thick, and pushes its inner disk radius, $ r_{in} = 2.026 ~ r_g $,
close to the marginally bound orbit, $ r_{mb} = 2 ~ r_g $, where the binding
energy is only $ |e_{in}| \approx 0.006 ~ c^2 $.
}
\end{figure}

\begin{figure}[t]
\vspace{8.0cm}
\includegraphics{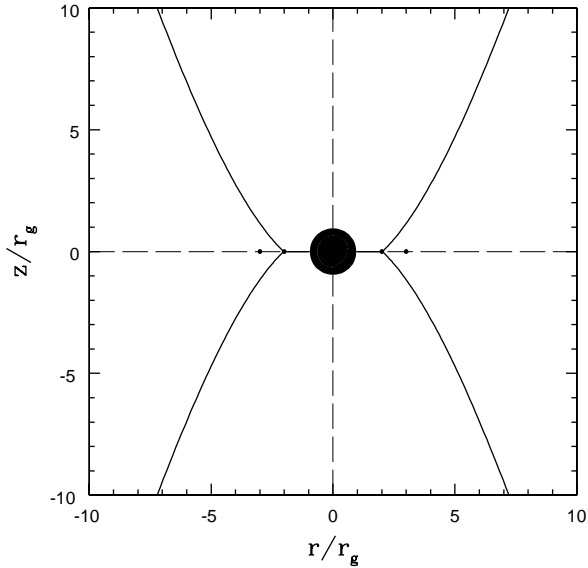}
\caption{
The inner parts of the toy disk model presented in Fig. 2.  The dots on the
equatorial axis indicate the location of the marginally bound orbit at
$ r_{mb} = 2 ~ r_g $, and the marginally stable orbit at $ r_{ms} = 3 ~ r_g $.
The disk becomes geometrically thin at $ r_{in} = 2.026 ~ r_g $, and cold
matter falls freely into the black hole at $ r < r_{in} $.  No thermal
energy is advected into the black hole in this model.
}
\end{figure}

As an example a toy disk model with $ r_{out} = 100 ~ r_g $ is shown
in Fig. 2 and Fig. 3.  The inner disk radius $ r_{in} = 2.026031... ~ r_g $
was obtained from the condition given with the eq. (20).  The eqs. (27)
were satisfied for 
$ a = 1.32048586889... $ and $ b = 0.000000442466655... $.

\begin{figure}[t]
\vspace{8.0cm}
\includegraphics{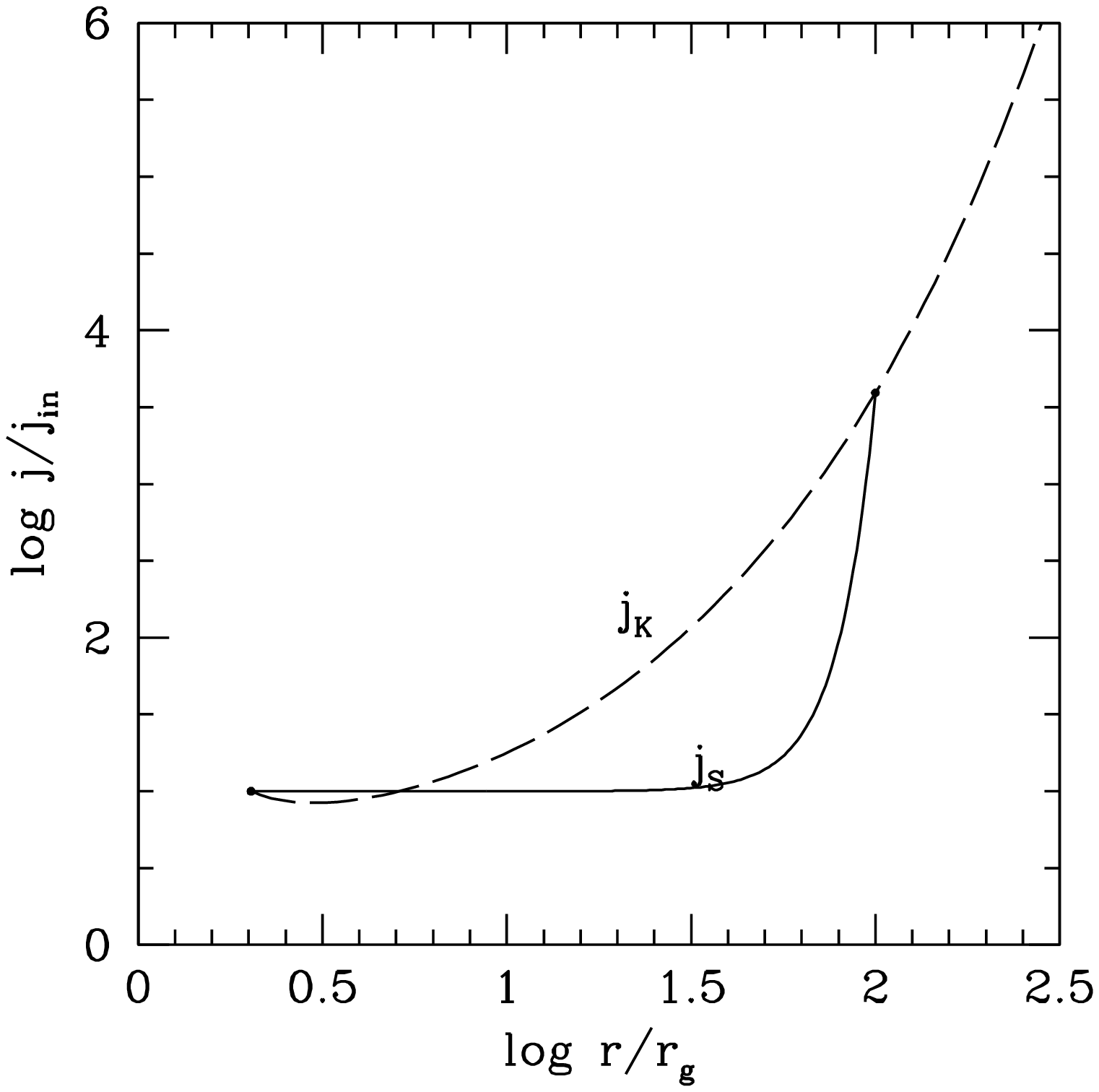}
\caption{
The variation of angular momentum with radius for the surface of
the thick disk presented in Figures 2 and 3 is shown with the solid line
($j_s$).  The variation of `Keplerian' angular momentum with radius is
shown with the dashed line ($j_K$).  Note that thick disk angular momentum is
almost constant for $ r < 10 ~ r_g $.
}
\end{figure}

\Section{Discussion}

The shape of our toy disk as shown in Fig. 2 is an artifact of the strong
assumptions made in this paper.  In particular, a rapid transition from
a thin disk to a thick disk at $ r_{out} $ is a direct consequence of the
assumption that 100\% of all energy dissipated locally is radiated away
locally for $ r > r_{out} $, while none is radiated away for $ r < r_{out} $.
In any realistic disk there will be partial radiation at all radii, and
the variation of the efficiency is likely to change gradually, with no
abrupt changes in the disk thickness.

While the detailed shape of the thick disk must be uncertain as long as we
have no quantitative understanding of disk viscosity, the formation of a thick
disk and pushing its inner radius towards the marginally bound orbit is a 
very general property.  It was noticed two decades ago with supercritical
accretion disks of Jaroszy\'nski et al. (1980) and Paczy\'nski and Wiita (1980).
In our toy model the inability of the disk to radiate energy dissipated within
$ r_{out} > r > r_{in} $ forces the disk to become thick, and pushes its inner
cusp towards $ r_{mb} $, lowering the efficiency with which rest mass is
converted into radiation.  The farther out the $ r_{out} $ is assumed to be
the closer $ r_{in} $ is pushed towards the $ r_{mb} $ in order to lower
the efficiency.

While we retain the term `advective' to describe our thick disk, it should
be stressed that as we require the disk to be thin at its inner cusp at
$ r_{in} $, there is no advection of any heat into the black hole, as all
enthalpy of the thick disk has been used up to press its inner radius 
towards the $ r_{mb} $.  However, the kinetic energy of thick disk matter
at $ r_{in} $ is much larger than is the kinetic energy of a thin disk,
which must have its inner radius located at $ r_{ms} $.  In the
pseudo-Newtonian potential the ratio of the two is $ v_{mb}^2/v_{ms}^2 = 8/3 $
(cf. eq. 2a).  

There is a somewhat paradoxical aspect of our toy model.  Some kind of
viscosity has to reduce angular momentum from $ j_{out} $ at the outer
disk boundary to $ j_{in} $ at its inner boundary, and this must generate
either heat or some other form of internal energy which puffs up the
disk, and it would seem that this energy has to be advected through
$ r_{in} $ into the black hole.  Figure 4 shows the distribution of
angular momentum for our thick disk model (solid line), and for the
`Keplerian' angular momentum given with the eq. (2c) (dashed line).
It is clear that thick disk angular momentum is almost constant for
$ r < 10 ~ r_g $, and that all dissipation takes place only in the outer
parts of the thick disk.  All the same the entropy must be higher at
$ r_{in} $ than it is at $ r_{out} $.  However, high entropy does not
have to imply high internal energy if the gas density is low at $ r_{in} $.

Obviously, our requirement for the disk to be thin at $ r_{in} $
is ad hoc, motivated by our desire to have as simple structure as possible.
Our toy model may require uninterestingly
low accretion rate to satisfy all the conditions that are imposed at 
its inner boundary: hydrostatic equilibrium, low geometrical thickness,
high entropy and low internal energy.  It is very likely that at the
accretion rate of any interest some of these conditions are broken.  It
is likely that the cusp at $ r_{in} $ opens up considerably, the speed
of sound is not negligible compared to rotational velocity, and the transonic
flow carries a non trivial amount of internal energy into the black hole
(Loska 1982). However, the conservation laws do not require the advected
thermal energy to be large, as demonstrated by our toy model.

A generic feature of any thick disk is the formation of a narrow funnel
along the rotation axis.  It is far from clear how realistic is the presence
of the funnel, as some instability might fill it in, and make the accretion
nearly spherical near the black hole.  On the other hand, if the funnel forms
and lasts, it may help collimate a jet-like outflow.  Unfortunately,
our lack of quantitative understanding of viscous processes in accretion
flows makes it impossible to prove what topology of solutions is realistic
under which conditions.  The main virtue of the toy model presented in this
paper is the set of assumptions that was made; this set is very different
from the assumptions which are used in
the recently booming industry of advection dominated accretion flows.  While
the assumptions adopted in this paper are ad hoc, so are the assumptions 
adopted in any thick disk models.  In particular, there is nothing `natural'
about the popular assumption that the accretion flow is self-similar.
Another popular assumption: constant `alpha' parameter is ad hoc as well.
It is useful to read the old paper by Galeev, Rossner and Vaiana (1979)
to realize that the very concept of the `alpha' parameter is ad hoc.
Disk properties which depend on any ad hoc
assumptions should not be taken seriously.

\Acknow{
It is a pleasure to acknowledge useful comments by the anonymous referee
which helped to improve the discussion of the toy model.  Dr. J.-P. Lasota
pointed out the errors in the original form of eq. (24) and the numerical
values of constants `a' and `b'; I am very grateful to him for his help.
}

\end{document}